\documentclass[11pt,a4paper]{article}
\usepackage{textcomp}
\usepackage{amssymb}
\usepackage{amsmath}
\usepackage{graphicx}
\usepackage{hyperref}
\usepackage{pgf}
\usepackage{graphicx}
\usepackage{epsfig}
\usepackage{graphics}
\usepackage{pstricks,pst-coil,pst-fill,pst-plot}
\usepackage{xypic}
\usepackage{enumerate}

\newtheorem{thm}{Theorem}[section]

\newtheorem{lemma}{Lemma}[section]
\newtheorem{cor}{Corollary}[section]

\newcommand{\be}[1]{\begin{equation}\label{#1}}
\newcommand{\ba}[1]{\begin{eqnarray}\label{#1}}
\newcommand{\ee}{\end{equation}}
\newcommand{\ea}{\end{eqnarray}}

\newcommand{\bra}[1]{\langle{#1}|}
\newcommand{\ket}[1]{|{#1}\rangle}

\let\qed=\cqfd

\def\sul{\sum\limits}
\def\pl{\prod\limits}

\def\build#1_#2^#3{\mathrel{
\mathop{\kern 0pt#1}\limits_{#2}^{#3}}}

\usepackage{euscript}

\def\la{\lambda}

\def\sul{\sum\limits}
\def\pl{\prod\limits}
\def\lt({\left(}
\def\rt){\right)}

\def\det{\operatorname{det}}

\def\la{\lambda}

\begin{document}
\begin{flushright}
\vphantom{LPENSL-TH-07/08}
\end{flushright}
\par \vskip .1in \noindent

\vspace{24pt}
\hfuzz 2pt
\begin{center}
\begin{LARGE}
{\bf 
\vspace*{1cm}
Partition function of the trigonometric SOS model with   reflecting end  }
\end{LARGE}

\vspace{50pt}

\begin{large}

{\bf G.~Filali}\footnote[1]{Universit\'e
de Cergy-Pontoise, LPTM UMR 8089 du CNRS,  2 av. Adolphe Chauvin, 95302 Cergy-Pontoise, France, e-mail: ghali.filali@u-cergy.fr},~~
{\bf N.~Kitanine}\footnote[2]{Universit\'e de Bourgogne,     				
Institut de Math\'ematiques de Bourgogne 		
UMR 5584 du CNRS,
9 av. Alain Savary - B.P. 47 870
21078 Dijon, France, e-mail: Nikolai.Kitanine@u-bourgogne.fr}

\end{large}

\vspace{80pt}

\centerline{\bf Abstract} \vspace{1cm}
\parbox{12cm}{\small We compute the partition function of the trigonometric SOS model with one reflecting end and domain wall type boundary conditions. We show that in this case, instead of a sum of determinants obtained by Rosengren for the SOS model on a square lattice without reflection, the partition function can be represented as a single Izergin determinant. This result is crucial for the study of the Bethe vectors of the spin chains with non-diagonal boundary terms.
}
\end{center}

\newpage

\section{Introduction}
The recent progress in the  study of the open XXZ spin chains with non-diagonal boundary terms \cite{Nep04,CaoLSW03,YanZ07, EssD05,EssD06}  permitted to apply the algebraic Bethe ansatz technique \cite{FadST79,Skl88} to this situation. It opens a possibility to investigate these models as well as some systems out of equilibrium such as ASEP. In particular there is a hope to compute the correlation functions using the algebraic Bethe ansatz technique \cite{KitMT99,KitMT00,KitKMNST07}. 

In this paper we consider  the  trigonometric SOS  model with a reflecting end. The reason to study such a particular case of general elliptic SOS model is the fact that  it is related to the XXZ chain with non-diagonal boundary conditions by a gauge transformation \cite{CaoLSW03,YanZ07} (similar to the gauge transformation permitting to reduce the XYZ chain to the SOS model). Our main goal is to compute the partition function for this system with {\it domain wall} type boundary conditions.

The domain wall boundary conditions for  exactly solvable models in two-dimensional statistical mechanics were introduced for the first time in the context of the calculation of the correlation functions \cite{Kor82} or, more precisely of computation of the scalar products and norms of the Bethe states. The partition function for the six-vertex model on a square $N\times N$ lattice with these conditions was obtained by Izergin \cite{Ize87} as a determinant of an
$N\times N$ matrix. This determinant representation is the corner stone for the computation of the correlation functions of quantum integrable models using algebraic Bethe ansatz \cite{IzeK84,KitMT99,KitMT00}.

For the open spin chains with diagonal boundary terms the same role is played by the six vertex model with one reflecting end and domain wall type boundary conditions on the other ones.  The partition functions of this system was studied by Tsuchiya \cite{Tsu98}. Once again the partition function was obtained as a determinant of a slightly more complicated $N\times N$ matrix. This representation led to a determinant representation for the scalar products and norms of the Bethe states \cite{YanZ07} and finally permitted to obtain the multiple integral representations for the correlation functions \cite{KitKMNST07,KitKMNST08}.

For the most general solution \cite{GhoZ94} non-diagonal solution of the reflection equation \cite{Che84} for the XXZ spin chain the algebraic Bethe ansatz using the gauge transformation which diagonalize the boundary matrix but transforms the usual trigonometric $R$ matrix (corresponding to the six-vertex model) into the dynamical trigonometric $R$ matrix (solution of the dynamical Yang-Baxter equation \cite{GerN84,Fel95,FelV06a,FelV06b}, corresponding to the trigonometric SOS model).

The SOS model with domain wall boundary condition was recently considered by different methods \cite{Ros09,PakRS08}. However even in the trigonometric case  the partition function  can not be expressed as a single determinant (it is obtained as a sum of determinants by Rosengren) and it makes the further steps much more complicated.

Here we add a reflecting end to the trigonometric SOS model. {\it A priori} it makes the result even more complicated. However we show in this paper that the corresponding partition  function once again can be written as a single determinant. In some sense the presence of a reflecting end permits to avoid some difficulties of the dynamical case. We hope that starting from this result it will be possible to compute the scalar products and the norms of the Bethe vectors for the most general open spin chains.

The paper is organized as follows. In the Section 2 we define the model and construct the corresponding dynamical reflection algebra. In the section 3 following \cite{Kor82} we establish the properties defining the partition function and  we show that there is a determinant solution for this functions.
\section{ SOS model and dynamical reflection equation.}

The SOS model is a two dimensional statistical mechanics lattice model which can be defined in terms of a {\it height function}. Every square of the lattice is characterized by a height $\theta$ and the its values for two adjacent squares differ by $\eta$. There are 6 possible face configurations

\[
\begin{array}{ccc}\begin{array}{ccc}
\vphantom{\sum\limits_1^2}\theta -\eta &\vline&\theta -2\eta\\
\hline
\vphantom{\sum\limits_1^2}\theta &\vline&\theta-\eta
\end{array}\qquad&\qquad
\begin{array}{ccc}
\vphantom{\sum\limits_1^2}\theta +\eta&\vline&\theta +2\eta\\
\hline
\vphantom{\sum\limits_1^2}\theta &\vline&\theta +\eta
\end{array}
\qquad&\qquad
\begin{array}{ccc}
\vphantom{\sum\limits_1^2}\theta -\eta&\vline&\theta \\
\hline
\vphantom{\sum\limits_1^2}\theta &\vline&\theta +\eta
\end{array}
\\
\bigskip
\\
\begin{array}{ccc}
\vphantom{\sum\limits_1^2}\theta +\eta&\vline&\theta \\
\hline
\vphantom{\sum\limits_1^2}\theta &\vline&\theta-\eta
\end{array}\qquad&\qquad
\begin{array}{ccc}
\vphantom{\sum\limits_1^2}\theta +\eta&\vline&\theta \\
\hline
\vphantom{\sum\limits_1^2}\theta &\vline&\theta +\eta
\end{array}
\qquad&\qquad
\begin{array}{ccc}
\vphantom{\sum\limits_1^2}\theta -\eta&\vline&\theta \\
\hline
\vphantom{\sum\limits_1^2}\theta &\vline&\theta -\eta
\end{array}
\\

\end{array}
\]

and the corresponding statistical weights $R^{a b}_{c d}$ can be written as an $R$ matrix acting in a tensor product of two two-dimensional spaces,
\begin{equation}
R(\lambda;\theta)=
\begin{pmatrix}
R^{+ +}_{+ +}(\lambda;\theta)&0&0&0\\
0 & R^{+ -}_{+ -}(\lambda;\theta)& R^{+ -}_{ - +}(\lambda;\theta) & 0\\
0 &R^{ - +}_{+ -}(\lambda;\theta) & R^{- +}_{- +}(\lambda;\theta) & 0\\
0 & 0 & 0 & R^{- -}_{- -}(\lambda;\theta)\\
\end{pmatrix}
\end{equation}
If the statistical weights do not depend on the height $\theta$ the model becomes equivalent to the six vertex model.

In general the model is exactly solvable if this $R$ matrix satisfies the Dynamical Yang-Baxter equation (DYBE)
\begin{align}
R_{12}(\lambda_{1}-\lambda_{2};\theta-\eta\sigma^{z}_{3})&R_{13}(\lambda_{1}-\lambda_{3};\theta)R_{23}(\lambda_{2}-\lambda_{3};\theta-\eta\sigma^{z}_{1})\nonumber \\=&R_{23}(\lambda_{2}-\lambda_{3};\theta)R_{13}(\lambda_{1}-\lambda_{3};\theta-\eta\sigma^{z}_{2})R_{12}(\lambda_{1}-\lambda_{2};\theta)
\end{align}

The most general solution of this equation can be written in terms of elliptic functions but here we consider only the trigonometric solution   with statistical weights
\begin{align}
R^{+ +}_{+ +}(\lambda;\theta)&=R^{- -}_{- -}(\lambda;\theta)=\sinh(\lambda+\eta) \nonumber\\
R^{+ -}_{+ -}(\lambda;\theta)&=R^{- +}_{ - +}(\lambda;-\theta)=\frac{\sinh\lambda\sinh(\theta-\eta)}{\sinh\theta} \\
R^{+ -}_{ - +}(\lambda;\theta)&=R^{- +}_{ + - }(\lambda;-\theta)= \frac{\sinh\eta\sinh(\theta-\lambda)}{\sinh\theta}\nonumber
\end{align}

We consider this model with a reflecting end, which means that each horizontal line makes a U-turn on the left side of the lattice. It produces two following configurations characterized by the weights  $\mathcal{K}_\pm^\pm (\lambda;\theta)$:

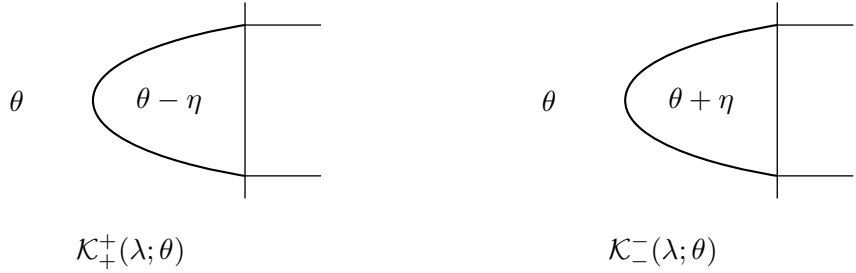
\begin{figure}[h]
\begin{center}
\setlength{\unitlength}{6cm}
\begin{pspicture}(12.5,4)
\psline[linewidth=0.5pt](4,1.5)(5,1.5)
\psline[linewidth=0.5pt](4,3.5)(5,3.5)
\psline[linewidth=0.5pt](4,1.2)(4,3.8)
\pscurve(4,1.5)(2,2.5)(4,3.5)
\rput(3,2.5){$\theta-\eta$}
\rput(1,2.5){$\theta$}
\rput(2.5,0.5){$\mathcal{K}_{+}^+(\lambda; \theta)$}

\psline[linewidth=0.5pt](11,1.5)(12,1.5)
\psline[linewidth=0.5pt](11,3.5)(12,3.5)
\psline[linewidth=0.5pt](11,1.2)(11,3.8)
\pscurve(11,1.5)(9,2.5)(11,3.5)
\rput(10,2.5){$\theta+\eta$}
\rput(8,2.5){$\theta$}
\rput(9.5,0.5){$\mathcal{K}_{- }^-(\lambda; \theta)$}
\end{pspicture}

\caption{Boundary configuration with external height $\theta$.}
\end{center}
\end{figure}
It's important to note that such reflecting end imposes a constant external height $\theta$ for the left side of the lattice.
Thus to preserve integrability the diagonal boundary matrix $\mathcal{K}(\lambda;\theta)$ should satisfy the usual reflection equation
\begin{align}
R_{12}(\lambda_{1}-\lambda_{2};\theta)&\mathcal{K}_{1}(\lambda_{1};\theta)R_{21}(\lambda_{1}+\lambda_{2};\theta)\mathcal{K}_{2}(\lambda_{2};\theta)\nonumber\\=&\mathcal{K}_{2}(\lambda_{2};\theta)R_{12}(\lambda_{1}+\lambda_{2};\theta)\mathcal{K}_{1}(\lambda_{1};\theta)R_{21}(\lambda_{1}-\lambda_{2};\theta)
\end{align}
The diagonal solution of this equation (which corresponds to a general solution for the six-vertex $R$-matrix after a gauge transformation \cite{YanZ07}) is
\begin{equation}
\mathcal{K}(\lambda;\theta)=
\begin{pmatrix}
\frac{\sinh(\theta+\zeta-\lambda)}{\sinh(\theta+\zeta+\lambda)}&0\\
0&\frac{\sinh(\zeta-\lambda)}{\sinh(\zeta+\lambda)}
\end{pmatrix}
\label{matrK}
\end{equation}
\\
This reflecting end lead to different parametrization of the weight if they are in the two different half the rows. Indeed, parametrization should respect row and line multiplication, and also some fundamental symmetry of the $R$ matrix (as the ice rule (6), that will be presented later). We can easily check that this convention lead to a well defined inhomogeneous model:
\\
\begin{figure}[ht]
\begin{center}
$R^{c-d, a-c}_{a-b, b-d}(\lambda_{j}+\xi_{i};a)$
\end{center}
\begin{center}
\includegraphics[width=2.5cm]{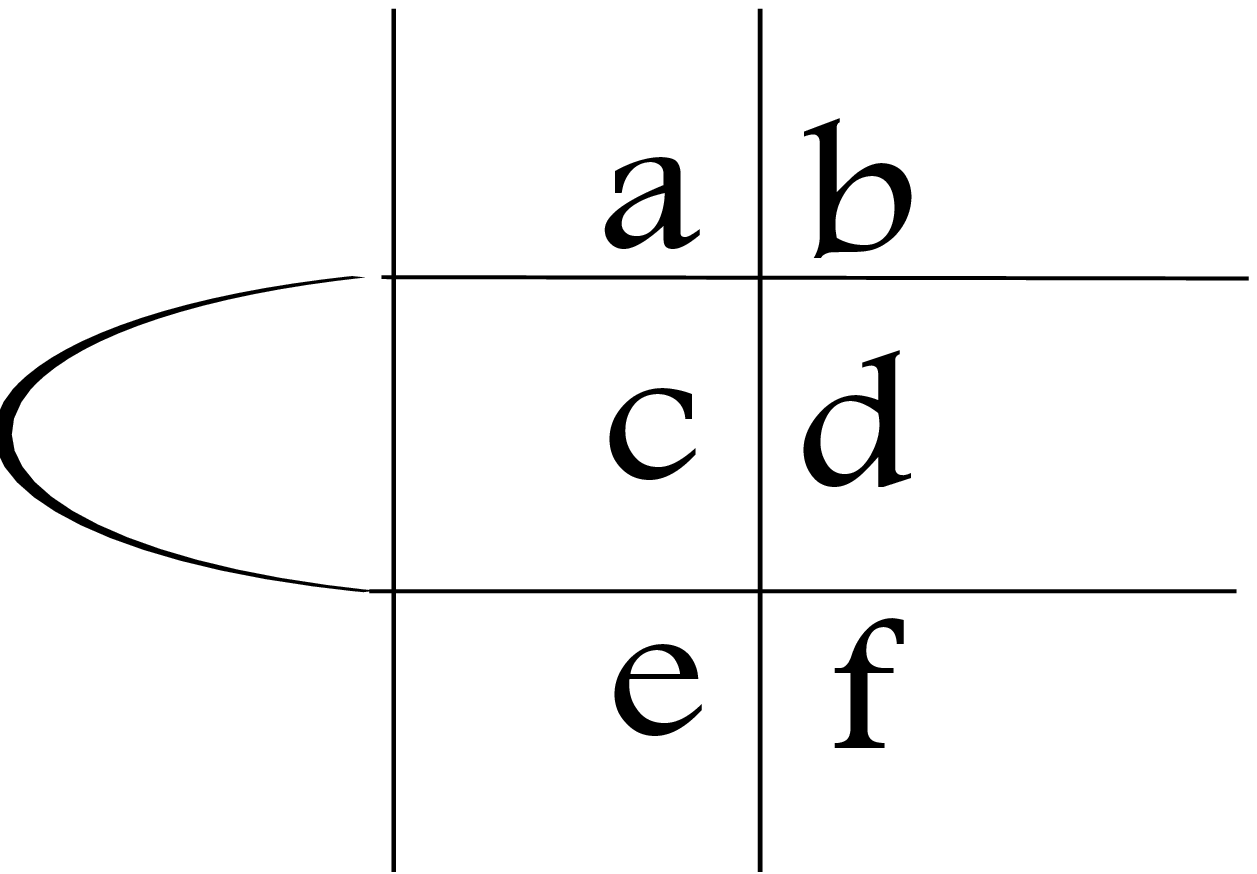}
\end{center}
\begin{center}
$R^{f-d, e-f}_{e-c, c-d}(\lambda_{j}-\xi_{i};e)$
\end{center}
\end{figure}
\vskip -0.5cm

The domain wall boundary conditions can be easily derived from the corresponding conditions for the six vertex model using the one to one correspondence between the configurations of the two models in the limit of the height-independent weights.
\begin{figure}[ht]
\begin{center}
\includegraphics[width=5.5cm]{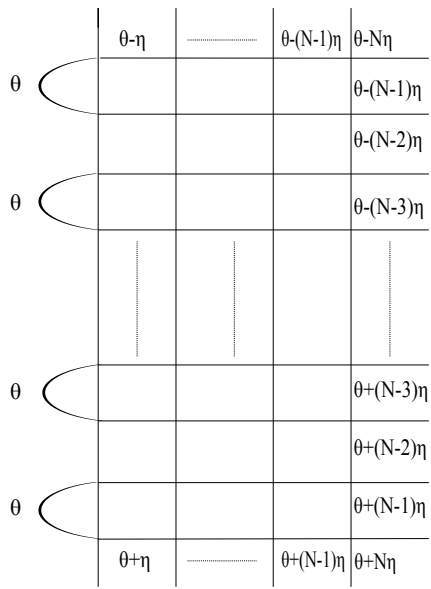}
\end{center}
\begin{center}
\caption{Domain Wall Boundary Conditions}
\end{center}
\end{figure}

One easily obtain the following boundary condition: the heights decrease from left to rights on the upper boundary, the heights grow from left to right on the lower boundary. As left external height is fixed these two conditions determine completely the configuration on the right boundary (heights decreasing in the upward direction).


In this paper we compute the partition function of this model. Once more time we would like to stress the point that while the model described above seems to be quite artificial it has a direct relation to the  spin chain with non-diagonal boundary terms and this partition function is a necessary step to compute the correlation functions for such chains.

\subsection{The bulk weights and their symmetry}
Prior to the computation of the partition function we need to establish some properties of the trigonometric SOS $R$ matrix and of the corresponding monodromy matrices.

This $R$ matrix  satisfies four important properties:
\\

{\bf 1.  Ice rule}
\medskip
\begin{equation}
[\sigma_{1}^{z}+\sigma_{2}^{z},R_{12}(\lambda;\theta)]=0
\end{equation}

This symmetry is responsible of the six vertex texture of the statistical weight: $R_{\alpha \beta}^{\mu \nu}=0$ unless $\alpha+\beta=\mu+\nu$. It is easy to see that this relation induce the following similar relations for the transposed $R$ matrix
\begin{equation}
[\sigma_{1}^{z}-\sigma_{2}^{z},R^{t_{1}}_{12}(\lambda;\theta)]=0
\end{equation}

{\bf 2. Unitarity}

\medskip

\begin{equation}
R_{12}(\lambda;\theta).R_{21}(-\lambda;\theta)=-\sinh(\lambda-\eta)\sinh(\lambda+\eta)Id
\end{equation}

\medskip

{\bf 3. Crossing Symmetry}

\medskip

The crossing relation for the dynamical $R$-matrix are not as simple as for the vertex type $R$ matrices, here we write it in the following compact form:
\begin{equation}
-\sigma_{1}^{y}:R_{12}^{t_{1}}(-\lambda-\eta;\theta+\eta\sigma_{1}^{z}):\sigma_{1}^{y}
\frac{\sinh(\theta-\eta\sigma_{2}^{z})}{\sinh\theta}
=R_{21}(\lambda;\theta)
\end{equation}
where we assume the following normal ordering: the $\sigma_{1}^{z}$ in the argument of the $R$ matrix (which does not commute with it) is always on the right of all other operators involved in the definition of $R$.

\subsection{Bulk monodromy matrix and double row monodromy matrix}

The bulk monodromy matrix  defined as
\begin{equation}
T_{0}(\lambda;\theta)=R_{01}(\lambda-\xi_{1};\theta-\eta\sum_{i=2}^{N}\sigma_{i}^{z})...R_{0N}(\lambda-\xi_{N};\theta)=
\begin{pmatrix}
A(\lambda;\theta)&B(\lambda;\theta)\\
C(\lambda;\theta)&D(\lambda;\theta)\\
\end{pmatrix}
\end{equation}
 satisfy  the dynamical Yang-Baxter algebra
\begin{align}
R_{12}(\lambda_{1}-\lambda_{2};\theta-\eta\sum_{i=1}^{N}\sigma^{z}_{i})&T_{1}(\lambda_{1};\theta)T_{2}(\lambda_{2};\theta-\eta\sigma^{z}_{1})\nonumber\\=&T_{2}(\lambda_{2};\theta)T_{1}(\lambda_{1};\theta-\eta\sigma^{z}_{2})R_{12}(\lambda_{1}-\lambda_{2};\theta)
\end{align}

To describe a reflecting end we introduce the double row monodromy matrix \cite{Skl88}
\begin{align}
\mathcal{T}(\lambda;\theta)\equiv&\begin{pmatrix}
\mathcal{A}(\lambda;\theta)&\mathcal{B}(\lambda;\theta)\\
\mathcal{C}(\lambda;\theta)&\mathcal{D}(\lambda;\theta)\\
\end{pmatrix}=T(\lambda;\theta)\mathcal{K}(\lambda;\theta)\widehat{T}(\lambda;\theta)\nonumber\\
=&R_{01}(\lambda-\xi_{1};\theta-\eta\sum_{i=2}^{N}\sigma_{i}^{z})...R_{0N}(\lambda-\xi_{N};\theta)\nonumber\\ \times&\mathcal{K}(\lambda;\theta)R_{N0}(\lambda+\xi_{N};\theta)...R_{10}(\lambda+\xi_{1};\theta-\eta\sum_{i=2}^{N}\sigma_{i}^{z})
\end{align}
As the $\mathcal{K}$ matrix  (\ref{matrK}) solves the (ordinary) reflection equation this double-row monodromy matrix satisfies  the following dynamical reflection equation
\begin{align}
R_{12}(\lambda_{1}-\lambda_{2};&\theta-\eta\sum_{i=1}^{N}\sigma^{z}_{i})\mathcal{T}_{1}(\lambda_{1};\theta)R_{21}(\lambda_{1}+\lambda_{2};\theta-\eta\sum_{i=1}^{N}\sigma^{z}_{i})\mathcal{T}_{2}(\lambda_{2};\theta)\nonumber\\
=\mathcal{T}_{2}(\lambda_{2};&\theta)R_{12}(\lambda_{1}+\lambda_{2};\theta-\eta\sum_{i=1}^{N}\sigma^{z}_{i})\mathcal{T}_{1}(\lambda_{1};\theta)R_{21}(\lambda_{1}-\lambda_{2};\theta-\eta\!\sum_{i=1}^{N}\sigma^{z}_{i})
\end{align}
This equation  contains the commutation  relations for the generators: $\mathcal{A}(\lambda;\theta)$, $\mathcal{B}(\lambda;\theta)$, $\mathcal{C}(\lambda;\theta)$ and $\mathcal{D}(\lambda;\theta)$, the only  one which is important for  the computation of the partition function is the relation for the $\mathcal{B}$ operators
\begin{equation}
\label{commutB}
\mathcal{B}(\lambda_{1};\theta)\mathcal{B}(\lambda_{2};\theta)=\mathcal{B}(\lambda_{2};\theta)\mathcal{B}(\lambda_{1};\theta)
\end{equation}
It is also important to establish some symmetries of the $\mathcal{B}$ operators.
Using  the crossing relation (9) and the Ice Rule for the transposed $R$ matrix (7) we  get:
\begin{align}
\widehat{T}(\lambda;\theta)\equiv & R_{N0}(\lambda+\xi_{N};\theta)...R_{10}(\lambda+\xi_{1};\theta-\eta\sum_{i=2}^{N}\sigma_{i}^{z})\nonumber\\
=&\gamma(\lambda)\sigma_{0}^{y}T^{t_{0}}(-\lambda-\eta;\theta+\eta\sigma_{0}^{z})\sigma_{0}^{y}\nonumber\\
&\qquad\qquad\quad
\frac{\sinh(\theta-\eta\sum_{i=1}^{N}\sigma_{i}^{z})}{\sinh(\theta)}\nonumber\\
=&\vphantom{\sum_{i=2}^{N}}\widehat{\gamma}(\lambda)T^{-1}(-\lambda;\theta)
\end{align}
with  normalization coefficients
\begin{equation}
\gamma(\lambda)=(-1)^{N},\widehat{\gamma}(\lambda)=(-1)^{N}\prod_{i=1}^{N}\sinh(\lambda+\xi_{i}-\eta)\sinh(\lambda+\xi_{i}+\eta)
\end{equation}
It implies for the $\mathcal{B}$ operators:
\begin{align}
\mathcal{B}(\lambda;\theta)=&\gamma(\lambda)\Big(\!\mathcal{K}^{-}_{-}B(\lambda;\theta)A(-\lambda-\eta;\theta+\eta)-\mathcal{K}^{+}_{+}A(\lambda;\theta)B(-\lambda-\eta;\theta-\eta)\!\Big)\nonumber\\
&\times \frac{\sinh(\theta-\eta\sum_{i=1}^{N}\sigma_{i}^{z})}{\sinh(\theta)}
\end{align}
And  using the dynamical Yang-Baxter algebra for the bulk monodromy matrix, this leads to the following symmetry of the $\mathcal{B}$ operators:
\begin{equation}
\mathcal{B}(-\lambda-\eta;\theta)=-\gamma(\lambda)\frac{\sinh(\lambda+\zeta)\sinh(2(\lambda+\eta))\sinh(\lambda+\zeta+\theta)}{\sinh(2\lambda)\sinh(\lambda-\zeta+\eta)\sinh(\lambda-\theta-\zeta+\eta)}\mathcal{B}(\lambda;\theta)
\label{crossB}
\end{equation}

\section{Partition Function}
The partition function of the SOS model introduced in the first section
can be written in terms of the double row monodromy matrix (this construction is parallel to the corresponding six-vertex partition function \cite{Tsu98})
\begin{align}
Z_{N,2N}(\{\lambda\},\{\xi\},\theta)=&\prod_{i=1}^{N} \uparrow_{\lambda_{i}}\prod_{j=1}^{N} \downarrow_{\xi_{j}}\prod_{i=1}^{N}   \mathcal{T}(\lambda_{i};\theta)
\prod_{i=1}^{N} \uparrow_{\xi_{i}}\prod_{j=1}^{N} \downarrow_{\lambda_{j}}\nonumber\\
=&
\bra{\bar{0}}\prod_{i=1}^{N}
\mathcal{B}(\lambda_{i};\theta)  \ket{0}\label{PartFun}
\end{align}
where $ \ket{0}$ is the state with all the spins up and $\ket{\bar{0}}$ is the state with all the spins down. We will follow the standard way to compute the partition function \cite{Kor82,Ize87}, first we establish a set of properties defining it in an unique way  and then we will propose a determinant formula which satisfies all these conditions

The partition function (\ref{PartFun}) satisfies the following  properties:
\begin{enumerate}[i)]
\item For each  parameter $\lambda_{i}$ the normalized partition function
\begin{align}
\tilde{Z}_{N,2N}(\{\lambda\},\{\xi\},\theta)&=\exp\left((2N+2)\sul_{i=1}^N\lambda_{i}\right)\nonumber\\
&\times\sinh(\theta+\zeta+\lambda_i)\sinh(\theta+\lambda_i)Z_{N,2N}(\{\lambda\},\{\xi\},\theta)\label{normPartFun}
\end{align}
is a polynomial of degree at most  $2N+2$ in
$e^{2\lambda_{i}}$.
This property follows immediately from the definition of the double row monodromy matrix.
\item For $N=1$ the partition function is  just a sum of two terms
\begin{multline}
Z_{1,2}(\lambda,\xi,\theta)=\frac{\sinh\eta\sinh(\theta-\eta)}{\sinh^2\theta}\\
\times\left(\frac{\sinh(\theta+\zeta-\lambda)}{\sinh(\theta+\zeta+\lambda)}\sinh(\lambda-\xi)\sinh(\theta+\lambda+\xi)\right.\\
\left.+\frac{\sinh(\zeta-\lambda)}{\sinh(\zeta+\lambda)}\sinh(\lambda+\xi)\sinh(\theta-\lambda+\xi)\right)
\end{multline}
as there are only two configurations possible.
\item $Z_{N,2N}(\{\lambda\},\{\xi\},\theta)$ is symmetric in $\lambda_{i}$.
This property follows from the commutation relation (\ref{commutB}) for the operators $\mathcal{B}$.
\item $Z_{N,2N}(\{\lambda\},\{\xi\},\theta)$ is symmetric in $\xi_{i}$.
This is a direct consequence of the Dynamical Yang-Baxter Equation. It is sufficient to insert $R_{i+1,i}(\xi_{i+1}-\xi_{i};\theta-\eta\sum_{j=i+2}^{N})$ in  (\ref{PartFun}) to get the symmetry for any elementary permutation ${\xi_{i}\leftrightarrow\xi_{i+1}}$.
\item Crossing symmetry.\\
For any  parameter $\lambda_{i}$ using the symmetry of the $\mathcal{B}$ operators (\ref{crossB}) we obtain the following relation
\begin{align}
Z_{N,2N}(-\lambda_{i}-\eta,\{\lambda\},&\{\xi\},\theta)=-\gamma(\lambda)\frac{\sinh(2(\lambda_{i}+\eta))\sinh(\lambda_{i}+\zeta)}{\sinh(2\lambda_{i})\sinh(\lambda_{i}-\zeta+\eta)}\nonumber\\
&\times \frac{\sinh(\lambda_{i}+\zeta+\theta)}{\sinh(\lambda_{i}-\theta-\zeta+\eta)}\, Z_{N.2N}(\lambda_{i},\{\lambda\},\{\xi\},\{\theta\})
\end{align}
\item Recursive relations.\\
  There are two points where we can easily establish recursive relations, fixing the configuration in the lower right or the upper right corner by setting  $\lambda_{1}=\xi_{1}$ or  $\lambda_{N}=-\xi_{1}$. It is easy to see that it leads to the following recursive relations
\begin{align}
Z_{N,2N}&\left. (\{\lambda\},\{\xi\},\{\theta\})\vphantom{\prod_{i=1}}\right|_{\la_1=\xi_1}=
\frac{\sinh\eta\sinh(\zeta-\lambda_1)}{\sinh(\zeta+\lambda_1)}
\nonumber\\
&\times\prod_{i=1}^{N}\sinh(\lambda_{i}+\xi_{1})\frac{\sinh(\theta+(N-2i)\eta)}{\sinh(\theta+(N-2i+1)\eta)}\nonumber\\
&\times\prod_{i=2}^{N}\sinh(\lambda_{1}-\xi_{i}+\eta)\sinh(\lambda_{1}+\xi_{i}+\eta)\sinh(\lambda_{i}-\xi_{1}+\eta)
\nonumber\\ &\times Z_{(N-1),2(N-1)}(\{\lambda\}_{2\dots N},\{\xi\}_{2\dots N},\{\theta\})
\end{align}
\begin{align}
Z_{N,2N}&\left. (\{\lambda\},\{\xi\},\{\theta\})\vphantom{\prod_{i=1}}\right|_{\la_N=-\xi_1}=
\frac{\sinh\eta\sinh(\theta+\zeta-\lambda_N)}{\sinh(\theta+\zeta+\lambda_N)}
\nonumber\\
&\times\prod_{i=1}^{N}\sinh(\lambda_{i}-\xi_{1})\frac{\sinh(\theta+(N-2i)\eta)}{\sinh(\theta+(N-2i+1)\eta)}\nonumber\\
&\times\prod_{i=2}^{N}\sinh(\lambda_{N}+\xi_{i}+\eta)\sinh(\lambda_{N}-\xi_{i}+\eta)\sinh(\lambda_{i-1}+\xi_{1}+\eta)\nonumber\\
&\times Z_{(N-1),2(N-1)}(\{\lambda\}_{1\dots N-1},\{\xi\}_{2\dots N},\{\theta\})
\end{align}
\end{enumerate}
\begin{lemma}
The set of conditions i)-vi) uniquely determine the partition function $Z_{N.2N}(\{\lambda\},\{\xi\},\{\theta\})$.
\end{lemma}
To prove this lemma it's sufficient to observe that the normalized partition function (\ref{normPartFun})
is a polynomial of degree at most $2N+2$ in each parameter $e^{2\la_i}$  defined in $4N$ points.
Indeed, due to the symmetries iii) and iv)  the recursion relations vi) can be established for any points $\la_i=\pm\xi_j$ . Due to the crossing symmetry v) similar recursion relations can be established in the points $\la_i=\mp\xi_j-\eta$. Hence we can prove by induction starting from the case $N=2$ that the partition function is uniquely determined. \qed

It means that if we find a function satisfying the above conditions it is the partition function.
\begin{thm}
The partition function of the trigonometric SOS model with reflecting end can be represented in the following form
\begin{align}
Z_{N,2N}&(\{\lambda\},\{\xi\},\{\theta\})=(-1)^N \det M_{i j}\prod_{i=1}^{N} \left(\frac{\sinh(\theta+\eta(N-2i))}{\sinh(\theta+\eta(N-2i+1))}\right)^{N-i+1}\nonumber\\
&\times\frac{\pl_{i,j=1}^{N}\sinh(\lambda_{i}+\xi_{j})\sinh(\lambda_{i}-\xi_{j})\sinh(\lambda_{i}+\xi_{j}+\eta)\sinh(\lambda_{i}-\xi_{j}+\eta)}{\pl_{1\leq i<j\leq N}\sinh(\xi_{j}+\xi_{i})\sinh(\xi_{j}-\xi_{i})\sinh(\lambda_{j}-\lambda_{i})\sinh(\lambda_{j}+\lambda_{i}+\eta)}
\end{align}
where the $N\times N $ matrix $M_{i j}$ can be expressed as a sum of two terms:
\begin{align}
M_{i,j}=\quad &\frac{\sinh(\theta+\zeta-\lambda_{i})}{\sinh(\theta+\zeta+\lambda_{i})}\, M_{i,j}^+ +
\frac{\sinh(\zeta-\lambda_{i})}{\sinh(\zeta+\lambda_{i})}\, M_{i,j}^-\nonumber \\
M_{i,j}^\pm=\quad &\frac{1}{\sinh(\lambda_{i}\mp \xi_{j}+\eta)}\left(\frac{1}{\sinh(\lambda_{i}\pm\xi_{j})}-\frac{\sinh(\theta\mp \eta)}{\sinh\theta\sinh(\lambda_{i}\pm\xi_{j}+\eta)}\right)
\label{PFresult}
\end{align}
\end{thm}
To prove the theorem it's sufficient to check the properties  i)-vi). The symmetries follow directly from the determinant structure and the $N=1$ case is evident. The recursion relations can be easily obtained  by  straightforward computation as  the determinant term is  reduced to a determinant of a $(N-1)\times(N-1)$ matrix.

It can be convenient  to express the matrix $M_{i j}$ in a slightly different form:
\begin{align}
M_{i,j}= &\frac{\sinh(\theta+\zeta+\xi_j)}{\sinh(\theta+\zeta+\lambda_{i})}\cdot\frac{\sinh(\zeta-\xi_j)}
{\sinh(\zeta+\lambda_{i})}\,\nonumber \\
\times &\frac{\sinh(2\lambda_i)\sinh\eta}{\sinh(\lambda_{i}- \xi_{j}+\eta)\sinh(\lambda_{i}+ \xi_{j}+\eta)\sinh(\lambda_{i}-\xi_{j})\sinh(\lambda_{i}+\xi_{j})}
\label{PFresult2}
\end{align}
In this form the result becomes very similar to the six vertex case \cite{Tsu98}.

\section*{Conclusion}

The main result of this paper is the fact  the partition functions of the SOS model with domain wall boundary conditions can be in some cases (namely if there is a reflecting end) expressed as a single determinant. This result opens some new possibilities for the study of the scalar products of the Bethe states and finally of the  correlation functions for the spin chains with non-diagonal boundary terms.

There are few other interesting questions arising from our result. First of all it should be quite easy to generalize it to the most general elliptic SOS model. The result should be once again a single determinant.  Also it is interesting to apply it in the particular ice-like point $\eta=i\frac \pi 3$ to study the three color model with a U-turn, following the analysis of Rosengren \cite{Ros09} and Kuperberg \cite{Kup02}.

\section*{Acknowledgments}

The authors are grateful to the LPTHE laboratory for hospitality which made this collaboration much easier. We would like to thank J.M. Maillet, V. Terras, J. Avan and V. Roubtsov for many interesting discussions.

\end{document}